\documentclass[fleqn,usenatbib]{mnras}

\usepackage[T1]{fontenc}

\DeclareRobustCommand{\VAN}[3]{#2}
\let\VANthebibliography\thebibliography
\def\thebibliography{\DeclareRobustCommand{\VAN}[3]{##3}\VANthebibliography}

\usepackage{subfig}
\usepackage{upgreek}
\usepackage{xcolor}

\usepackage{graphicx}	
\usepackage{amsmath}	

\title[Notes on PBH origin for thermal GRBs]{Notes on primordial black hole origin for thermal gamma-ray bursts}

\author[T. McMaken]{
Tyler McMaken$^{1}$\thanks{E-mail: tyler.mcmaken@colorado.edu}
\\
$^{1}$JILA and Department of Physics, University of Colorado, Boulder, Colorado 80309, USA
}

\pubyear{2021}

\usepackage{newtxtext,newtxmath}

\begin{document}
\label{firstpage}
\pagerange{\pageref{firstpage}--\pageref{lastpage}}
\maketitle

\begin{abstract}
Recently, an alleged plausible astrophysical scenario was proposed for the production of observed thermal gamma-ray bursts, via Hawking radiation emitted from a primordial black hole (PBH) freely falling into a more massive black hole. Here the implausibility of that scenario is demonstrated, and the key flaws in that paper's calculations and assumptions are elucidated through a discussion of some common misconceptions concerning black holes and general relativity. In particular, the predicted radiance observed from Earth is found to be orders of magnitude lower than what any instrument could detect, and the PBH-BH merger signature would be completely overwhelmed by the background Hawking signature from free PBHs.
\end{abstract}


\section{Introduction}\label{sec:introduction}

Gamma-ray bursts (GRBs) are some of the most energetic events observed in the Universe. Such events are typically characterized by non-thermal spectra, and despite the diversity in their light curves and the uncertainty in their exact emission mechanisms, the sources for GRBs are almost unanimously agreed upon to fall under two classes, based on the GRB's duration.

Short GRBs, with a duration of tenths of a second, are associated with kilonovae produced from compact binary mergers. Such a view has been confirmed both computationally and observationally \citep{nak07,met10,tan13,ber13}, notably with the detection of the short GRB 170817A alongside LIGO's and VIRGO's detection of gravitational wave GW170817 from a neutron star merger event \citep{abb17}. On the other hand, the progenitors of long GRBs, with a duration of tens of seconds, are massive stars undergoing core-collapse \citep{mac01,woo06}. This collapsar model has been confirmed with numerous coincident observations, including GRB 980425 \& SN 1998bw \citep{kul98}, GRB 030329 \& SN 2003dh \citep{pao03}, and GRB 060218 \& SN 2006aj \citep{sol06}.

Because of the diversity in the population of GRBs and the continued speculation surrounding the exact details of the emission mechanisms from the above-mentioned progenitors, some have proposed alternative exotic GRB sources. Most recently, \citet{bar21} has claimed that the Hawking radiation from an atom-sized primordial black hole (PBH) falling into a larger black hole would produce a thermal spectrum consistent with that of a GRB. According to his model, as the PBH falls sufficiently close to the horizon, the Hawking temperature undergoes a significant Lorentz boost and leads to an energetic, highly collimated beam of blackbody radiation that reaches Earth.

Aside from the fact that pure thermal spectra are only observed for a rare handful of GRBs that are already well-explained by the fireball model \citep{ghi13}, the work of \citet{bar21} contains several misleading claims, erroneous calculations, and nonphysical assumptions, most notably in the use of special relativistic equations to describe Schwarzschild near-horizon behavior and in the choice of ${1.1\times10^{17}}$ for the PBH's initial Lorentz factor $\gamma$. Thus, a follow-up analysis of the work is warranted.

This work will not consider the (un)likelihood of the existence of PBHs as a source for dark matter, as the subject has already been considered at length in the literature (see, e.g.\ \citet{vil21} for a review). Instead, this work is concerned solely with the viability of the detection of such objects if they were to fall into a black hole. Section~\ref{sec:calculation} outlines a rectified approach to the problem considered by Barco, yielding a vastly different light curve than what he has presented. Then, Section~\ref{sec:misconceptions} enumerates specific misconceptions that one may glean from Barco's work, and Section~\ref{sec:discussion} concludes the paper.

\section{Calculation}\label{sec:calculation}

In what follows, geometrized units are used where $c=G=\hbar=k_B=1$. The metric signature is ${(-+++)}$, and with the exception of the subscript $\omega$, all Greek indices are tensorial spacetime indices. Section~\ref{subsec:problemsetup} details the setup of the problem considered by \citet{bar21}, the calculation itself (different from Barco's ad hoc approach) is performed in Section~\ref{subsec:radiancecalculation}, and numerical results are presented in Section~\ref{subsec:numericalresults}.

\subsection{Problem setup}\label{subsec:problemsetup}

Consider a Schwarzschild black hole with mass $M_\bullet$, described by the line element
\begin{equation}
    ds^2=-\Delta dt^2+\Delta^{-1}dr^2+r^2\left(d\vartheta^2+\sin^2\!\vartheta d\varphi^2\right),
\end{equation}

\noindent where ${\Delta(r)\equiv1-2M_\bullet/r}$ is the horizon function.
To this spacetime add a small, spherical object with mass $M_\text{e}\ll M_\bullet$ and radius $2M_\text{e}$, located at ${r=r_\text{e}}$ (where the subscript e stands for ``emitter,'' since it will be emitting Hawking radiation). Without loss of generality, the object will be placed at the pole ${\vartheta=0}$ to remove any dependence on the azimuthal angle $\varphi$. This object will model a PBH, which will be assumed (as is standard) to be in free fall from rest at infinity (this assumption will be revisited in Section~\ref{subsec:vacuum} and will be found to be approximately true for physical scenarios). Quantitatively, this free fall condition implies that the object's four-velocity ${u_\text{e}^\mu\equiv dx_\text{e}^\mu/d\tau}$ in the static coordinate frame ${(t,r,\vartheta,\varphi)}$ is \citep{mis73}:
\begin{equation}\label{eq:ue}
    u_\text{e}^\mu=\left(\frac{E_s}{\Delta(r_\text{e})},\quad-\sqrt{E_s^2-\left(1+\frac{L_s^2}{r_\text{e}^2}\right)\Delta(r_\text{e})},\quad-\frac{L_s}{r_\text{e}^2},\quad0\right).
\end{equation}

\noindent The specific energy $E_s$ and specific angular momentum $L_s$ will be left arbitrary for now, though it should be noted that the condition of free fall from rest at infinity implies that ${E_s=1}$.

Further, assume that the PBH emits blackbody radiation isotropically from its surface with a constant Hawking temperature $T'$ in its own rest frame. Each photon from the PBH begins with an angular frequency ${\omega'(r_\text{e})}$ corresponding to the emitter's rest frame and propagates outward until reaching an observer on Earth, located a distance $r_o$ from the black hole. Here and elsewhere, primed quantities are measured in the free fall frame (i.e.\ the rest frame of the emitter), and unprimed quantities are measured in the static frame (i.e.\ the rest frame of the observer on Earth).

The observer on Earth is assumed to be static, with a four-velocity
\begin{equation}\label{eq:uo}
    u_\text{o}^\mu=\left(\frac{1}{\sqrt{\Delta(r_\text{o})}},\quad0,\quad0,\quad0\right).
\end{equation}

\noindent Any other four-velocity can be inserted here if one wishes to consider a PBH-BH system moving with respect to Earth, but since other galaxies are only redshifting and receding from Earth, such a choice would only act to decrease the amount of observed radiation.

The goal is then to calculate the spectral irradiance from the PBH emitter detected by an observer on Earth.

\subsection{Radiance calculation}\label{subsec:radiancecalculation}

Though radiation emitted just above the horizon may be modified by absorption and emission from a black hole's accretion disc, the simplified setup of this problem assumes that no radiative transfer is required---the spectral radiance from the emitter simply needs to be ray-traced to the observer on Earth along a null geodesic bundle.

The spectral radiance $I_{\omega'}$ (also called the specific intensity) of a photon with angular frequency $\omega'$ is defined by:
\begin{equation}\label{eq:Idef}
    \omega'd\mathcal{N}\equiv I_{\omega'}\cos\theta'\ d\omega'\ d\Omega'\ dA'\ d\tau',
\end{equation}

\noindent where $d\mathcal{N}$ is the number of photons emitted from an area element $dA'$ in the emitter's proper time $d\tau'$ through the solid angle $d\Omega'$, and $\theta'$ is the angle at which each photon is emitted to the normal of the area element \citep{ryb85,yos19}.

For a blackbody emitter, the spectral radiance is given by Planck's law:
\begin{equation}
    I_{\omega'}=\frac{\omega'^3}{2\pi^2\left(\text{e}^{\omega'/T'}-1\right)},
\end{equation}

\noindent where the Hawking temperature $T'$ emitted in the PBH's rest frame is assumed to be constant over the duration of the free fall event:
\begin{equation}
    T'=\frac{1}{8\pi M_\text{e}}.
\end{equation}

As Barco notes, introducing a spin for the PBH (i.e.\ using Kerr instead of Schwarzschild) will slightly modify this formula, but doing so can only decrease the PBH's Hawking temperature.

To calculate the spectral radiance ${I_\omega(r_\text{o})}$ observed from Earth, a key insight from analyzing how equation~(\ref{eq:Idef}) behaves under transformations is that the occupation number ${f\propto{I_\omega}/\omega^3}$ is Lorentz-invariant \citep{mis73}. Thus, one can write:
\begin{equation}\label{eq:invariant}
    I_\omega(r_\text{o})=g^3I_{\omega'}(r_\text{e}),
\end{equation}

\noindent where ${g\equiv\omega(r_\text{o})/\omega'(r_\text{e})}$ is the blueshift factor, greater than 1 for blueshifted photons and less than 1 for redshifted photons (note that $g$ is distinct from the astronomer's redshift factor ${z\equiv g^{-1}-1}$).

Now all that needs to be done is to calculate the frequencies $\omega(r_\text{o})$ and $\omega'(r_\text{e})$, representing the time components of the photon four-momentum measured in the frame of and evaluated at the position of the observer and the emitter, respectively. The angular frequency in either frame is given by \citep{mis73}:
\begin{equation}\label{eq:omega}
    \omega\equiv-u_\mu k^\mu,
\end{equation}

\noindent where the four-velocity ${u_\mu=g_{\mu\nu}u^\nu}$ is given by equation~(\ref{eq:ue}) or (\ref{eq:uo}), and the photon's coordinate frame four-momentum ${k^\mu\equiv dx^\mu/d\lambda}$ (normalized to its frequency at infinity) is
\begin{equation}\label{eq:k}
    k^\mu=\left(\frac{1}{\Delta},\quad\sqrt{1-\frac{b^2}{r^2}\Delta},\quad-\frac{b}{r^2},\quad0\right),
\end{equation}

\noindent for the impact parameter $b$. The choice of $b$ will uniquely determine the specific photon path connecting the emitter to the observer. Equivalently, the photon path can be constrained by specifying the position $r_\text{e}$ of the emitter and the angle $\theta_\text{e}$ at which the photon is emitted with respect to ${\hat{r}}$. The relation between $b$ and $\theta_\text{e}$ can be found by transforming $k^\mu$ into the static orthonormal tetrad frame and solving ${\tan\theta_\text{e}=k^{(2)}_\text{e}/k^{(1)}_\text{e}}$, with the result:
\begin{equation}\label{eq:b}
    b=\frac{r_\text{e}\sin\theta_\text{e}}{\sqrt{\Delta(r_\text{e})}}.
\end{equation}

Combining equations~(\ref{eq:ue}), (\ref{eq:uo}), (\ref{eq:omega}), (\ref{eq:k}), and (\ref{eq:b}), the blueshift factor becomes
\begin{equation}
    g=\sqrt{\frac{\Delta(r_\text{e})}{\Delta(r_\text{o})}}\left(\frac{E_s}{\sqrt{\Delta(r_\text{e})}}+\cos\theta_\text{e}\sqrt{\frac{E_s^2}{\Delta(r_\text{e})}-1-\frac{L_s^2}{r_\text{e}^2}}-\frac{L_s}{r_\text{e}}\sin\theta_\text{e}\right)^{-1}.
\end{equation}

\noindent The observed spectral radiance then becomes
\begin{equation}\label{eq:Io}
    I_\omega(r_\text{o})=\frac{\omega(r_\text{o})^3}{2\pi^2\left(\text{e}^{\omega(r_\text{o})/T_\text{eff}}-1\right)},
\end{equation}

\noindent with the effective temperature
\begin{equation}\label{eq:Teff}
    T_\text{eff}=gT'.
\end{equation}

\noindent Any other radiometric quantities, such as the spectral irradiance ${F_\omega=\int I_\omega d\Omega}$ or the irradiance $F=\int F_\omega d\omega$, can then be derived from equation~(\ref{eq:Io}).

\subsection{Numerical results}\label{subsec:numericalresults}

The effective temperature obtained in the Section~\ref{subsec:radiancecalculation}, equation~(\ref{eq:Teff}) looks quite different than the analogous temperature calculated by \citet{bar21} in his equation~(18). Part of the reason for the differences is that the derivation here uses the emission angle $\theta_\text{e}$ measured in the observer's frame, whereas Barco uses the angle $\theta'_\text{e}$ measured in the emitter's frame. Using $\theta'_\text{e}$ makes the algebra needlessly more complicated; for example, the relationship between $b$ and $\theta'_\text{e}$ analogous to equation~(\ref{eq:b}) will include extra $L_s$-dependent terms. Barco's analysis neglects many of these terms, and his assumption that ${v^{(r)}\gg v^{(\varphi)}}$ should actually remove any $L_s$-dependence from his final expression.

\begin{figure}
	\includegraphics[width=\columnwidth]{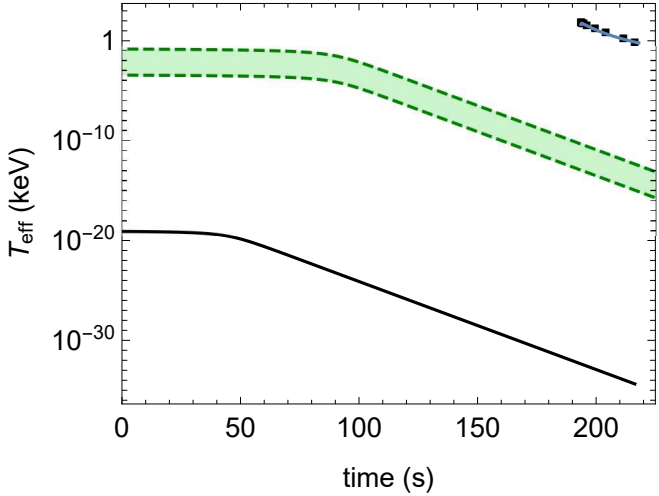}
    \caption{Effective temperature $T_\text{eff}$ as a function of the coordinate time $t$ as a PBH falls from a distance of $\sim$20 Gm above the horizon to a distance of $\sim$1 $\upmu$m above the horizon. The green shaded region shows $T_\text{eff}$, calculated from equations~(\ref{eq:ue}) and (\ref{eq:Teff}), for a realistic set of parameters (see Sec.~\ref{subsec:vacuum}) over the allowed PBH mass range ($10^{-11}\sim10^{-14}M_\odot$). For comparison, the black curve shows the value of $T_\text{eff}$ assuming the parameters used by \citet{bar21} (${M_\bullet=5\times10^5M_\odot}$, ${M_\text{e}=2.5\times10^{-13}M_\odot}$, ${L_s\approx2.23\times10^{-13}}$, and ${E_s=1.1\times10^{17}}$), and the black squares and blue curve show the analogous values presented by Barco.}
    \label{fig:cooling}
\end{figure}

What, then, will a radiating PBH actually look like as it falls into a more massive black hole? The spectrum will appear as a blackbody, in accordance with equation~(\ref{eq:Io}). The observed temperature of this blackbody will be given by equation~(\ref{eq:Teff}) and will cool as the PBH approaches the horizon, but at a different rate and a much lower overall magnitude than that predicted by Barco. Fig.~\ref{fig:cooling} shows the comparison between the cooling behavior found here and that found by \citet{bar21} (cf.~Fig.~4 of that work).

As can be seen from Fig.~\ref{fig:cooling}, the observed temperature of the PBH will actually be much lower than that calculated by Barco. The choice of constants used for the black curve in this plot is the same as that chosen by Barco (in particular, ${E_s=1.1\times10^{17}}$), while a more realistic, physically motivated set of constants are used for the green shaded region; a discussion of the physical validity of those constants is deferred to Sec.~\ref{subsec:vacuum}.

The effective temperature depends inversely on both $E_s$ (with the green region approximately at the lower bound for $E_s$) and $M_\text{e}$ (the PBH would need a mass of ${M_\text{e}\sim10^{-30}M_\odot}$ to match Barco's values in the observable regime). The choice of $L_s$ only changes $T_\text{eff}$ by less than an order of magnitude, and the choice of $M_\bullet$ only changes the scaling of the Schwarzschild time and distances. Also, note that Barco's equation (20) (used to calculate the blue curve in Fig.~\ref{fig:cooling}) contains a typo and should read ``$t_n/t_0$'' in the denominator instead of ``$t/t_n$''. Barco also excludes a necessary factor of $\delta$ in the exponent when performing his broken power law fit \citep{ryd04}.

The dependence of the emission from the infalling PBH on the observer's time $t$ is found by integrating ${u^t_\text{e}/u^r_\text{e}}$ from equation~(\ref{eq:ue}) along the null geodesic connecting the observer and emitter. The polar coordinate $\theta_\text{e}$ will also change as the PBH approaches the horizon. Though it is not stated explicitly, if Barco did use the same constant value for this angle for different emitter radii, he would essentially be relocating the position of Earth for every new position $r_\text{e}$ as the emitter falls in, so that Earth is always in the optimal spot to receive maximum radiation.

As a final comment, note that Fig.~\ref{fig:cooling} traces back the PBH much farther from the horizon than what Barco had presented. Doing so reveals that for the given choices of parameters, the observed radiation would have lasted much longer than Barco's reported 22.7 seconds---Barco considers only the final microns of the PBH's descent, though much more radiation will actually be observed when the PBH is farther from the horizon. In fact, the most radiation would actually be observed in the asymptotic regime before the PBH even reaches the black hole. The PBH dark matter scenario would thus predict a free particle Hawking background many orders of magnitude higher than the radiation signature from PBH-BH mergers.

\section{Misconceptions}\label{sec:misconceptions}

To explain the key differences in these calculations and the assumptions that lead to these differences, the discussion that follows will be structured around three main misconceptions about general relativity that pervade Barco's work.

\subsection{Special relativity isn't an add-on to general relativity}

The first misconception concerns the applicability of formulas from special relativity and general relativity. The equations of special relativity apply to flat spacetimes, whereas general relativity encompasses all spacetimes. Though special relativistic effects like kinematical boosting may be relevant for fast objects in non-flat spacetimes, these effects do not need to be added in separately, since they are already encompassed by the equations of general relativity.

As an example, consider the blueshift factor $g$ appearing in equation~(\ref{eq:invariant}). This factor can be expressed in a form that reveals two independent effects \citep{yos19}:
\begin{equation}\label{eq:g_grav_dopp}
    g\equiv\frac{\omega(r_\text{o})}{\omega'(r_\text{e})}=\frac{\omega(r_\text{o})}{\omega(r_\text{e})}\frac{\omega(r_\text{e})}{\omega'(r_\text{e})}=\left(\sqrt{\frac{\Delta(r_\text{e})}{\Delta(r_\text{o})}}\right)\left(\frac{1}{\gamma\left(1+\beta\cos\theta_\text{e}\right)}\right),
\end{equation}

\noindent which is valid when ${L_s=0}$. The first term in parentheses on the right hand side of equation~(\ref{eq:g_grav_dopp}) is the gravitational redshift between the near-horizon emitter and the observer, whereas the second term is associated with the Doppler effect from the relative motion between the emitter and the observer. Both of these effects are implicitly included in the calculations of Section~\ref{sec:calculation}; no additional relativistic formulas are needed.

In contrast, Barco adds in factors from relevant effects ad hoc, and more problematically, uses some equations that are only valid in special relativity. One example, cast into the notation of this paper, is his equation~(11):
\begin{equation}
    T=T'\gamma\left(1-\beta\cos\theta_\text{e}\right)
\end{equation}

\noindent(and note that the Lorentz factor $\gamma$ given by equation~(17) of \citet{bar21} contains a typo and should read $E_s^{-2}$ instead of $E_s^2$). This equation correctly models the deboosting of radiation from a blackbody emitter travelling away from an observer, but it is only applicable in flat spacetimes \citep{hen68}. For an emitter mere microns away from the horizon of an intermediate mass black hole, spacetime is certainly not flat. Deboosting effects will still be present, but they will not take the same form as in the simple flat case and instead will naturally appear in the full general relativistic calculation.

\begin{figure*}
    \centering
    \subfloat[\centering]{{\includegraphics[width=0.8\columnwidth]{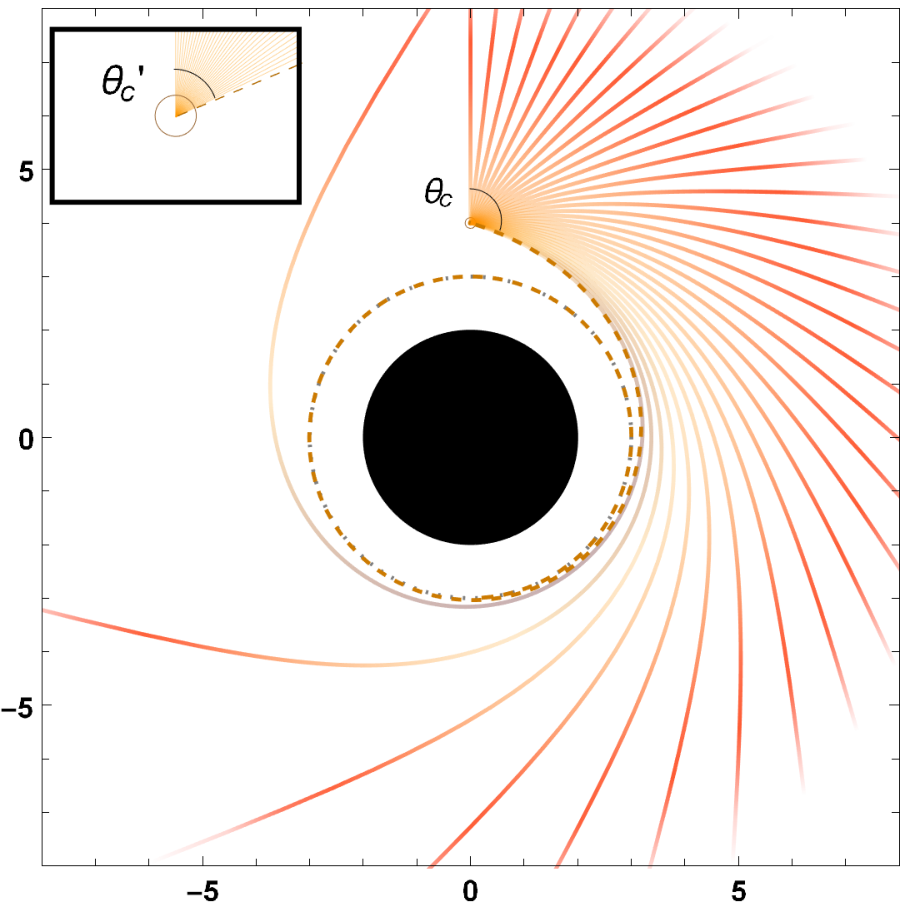} }}%
    \qquad
    \subfloat[\centering]{{\includegraphics[width=0.8\columnwidth]{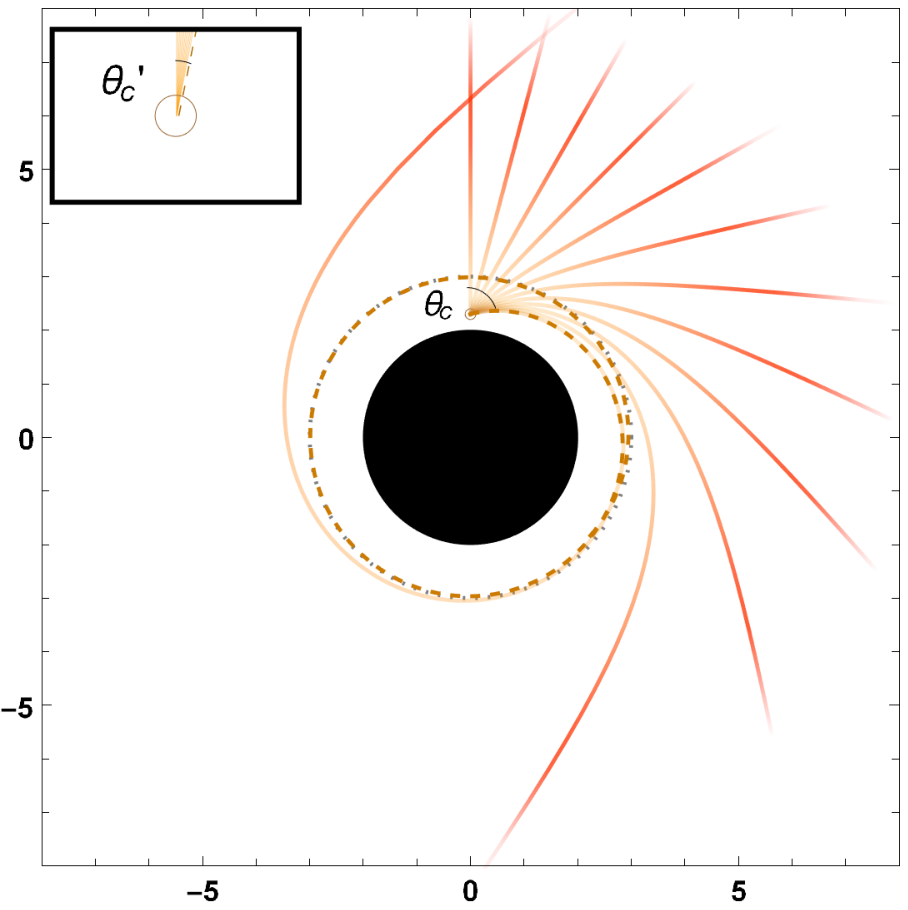} }}%
    \caption{Null geodesics emanating from an emitter (small, brown circle) at a radius (a) ${r_\text{e}=4M_\bullet}$ and (b) ${r_\text{e}=2.3M_\bullet}$. The geodesics are equally spaced in the free fall frame with a separation of $2^\circ$ for (a) and $1^\circ$ for (b), and for simplicity, only the right half of the set of geodesics that can reach infinity are shown. The insets in the upper left show a close-up of the geodesics in the emitter's local tetrad frame. The color shows the degree of redshift along each geodesic, and the dashed curve shows the geodesic that asymptotes to the photon sphere. The black disc shows the portion of the black hole within the horizon (${r\le2M_\bullet}$).}
    \label{fig:plotcones}
\end{figure*}

\subsection{Black holes don't enhance/collimate near-horizon light}

As an external observer views an emitter approaching the horizon of a black hole, the radiation from that emitter is exponentially dimmed and redshifted until it merges with the extremely cold Hawking radiation noise from the black hole itself. Why, then, does Barco argue that the radiation from the PBH is enhanced as it approaches the horizon until it forms an ``extremely collimated beam''?

Part of the confusion comes from the concept of an escape cone. As an isotropic emitter falls into a black hole, the amount of radiation that is able to escape to infinity decreases until the emitter passes the horizon, after which all radiation (even photons travelling directly outwards in the emitter's rest frame) will remain trapped in the black hole. The escape cone is the solid angle of radiation from the near-horizon emitter that is able to escape to infinity. This cone is not a concentrated enhancement of the radiation; it is simply an ever-decreasing selection of a portion of the isotropic radiation.

In fact, the radiation from the escape cone is not collimated in the slightest. As shown in Fig.~\ref{fig:plotcones}, for any external emitter near the black hole, light from the escape cone will diffusely spread across the entire sky. Only a vanishingly small portion of this radiation will reach Earth. The light may appear collimated in the emitter's local free fall frame (the upper left insets of Fig.~\ref{fig:plotcones}), but when transforming back to the observer's frame, it is clear that the light does not form a concentrated beam.

Once again, Barco misapplies special relativity here, in the assumption that the least deboosted radiation will come from the edge of the escape cone (at a critical angle $\theta_c'$) and travel straight back to Earth. In actuality, photons emitted at the angle $\theta_c'$ will circle around the black hole as they asymptotically approach the photon sphere (this path is shown by the dashed curves in Fig.~\ref{fig:plotcones}) and eventually veer off to infinity. Such a choice for the emission angle would be an extremely poor choice if one were using the correct equations from general relativity, since a vanishingly small number of the ray-traced photons from this angle would actually reach Earth after their endless traverse along the photon sphere.

\subsection{Black holes are not cosmic vacuum cleaners}\label{subsec:vacuum}

The final and perhaps most widespread misconception about a black hole is that it is a sort of cosmic ``vacuum cleaner'' that sucks in everything even remotely close to it at near-light speeds. But the truth is that black holes have no special power beyond that of any other gravitating mass. If the Sun were suddenly replaced by a 1 solar mass black hole, the gravitational dynamics of the solar system would remain completely unchanged.

How does this misconception come into play in Barco's model? First, note that it is extremely improbable for an object with a random trajectory to be captured by a black hole. The black hole will not suck in everything nearby; since PBHs are essentially collisionless, any merger must be the result of a direct head-on collision with an extremely compact region of space. Even objects already in a black hole's accretion disc will rarely fall within the innermost stable circular orbit (ISCO) and dive below the horizon.

To make this first note more quantitative, consider the capture rate of PBHs by more massive black holes given the currently observed local dark matter density of ${\rho_\text{DM}\sim0.5}$ GeV/cm$^3$ \citep{rea14}. The capture rate $\mathcal{F}$ can be estimated by integrating a Maxwellian dark matter distribution over the parameter space of $E_s$ and $L_s$ that would lead to capture within a radius $R$, with $E_s$ up to the characteristic scale $(1/3)\bar{v}^2$ and $L_s$ constrained by equation~(\ref{eq:Ls}) below:
\begin{equation}
    \mathcal{F}=\frac{\Omega_\text{PBH}}{\Omega_\text{DM}}\sqrt{6\pi}\frac{\rho_\text{PBH}}{m_\text{PBH}}\frac{2M_\bullet R}{\bar{v}(1-2M_\bullet/R)}
\end{equation}

\noindent\citep{kou08,cap13}, where the velocity dispersion $\bar{v}$ is taken to be 7 km/s. Even in the extremely unlikely scenario that $m_\text{PBH}$ were small enough for radiation to be detectable from the few black holes within a kiloparsec from Earth, and all dark matter were composed of PBHs (i.e.\ ${\Omega_\text{PBH}=\Omega_\text{DM}}$), and all PBHs that reach within a black hole's ISCO were captured, the estimated capture rate would be
\begin{equation}
    \mathcal{F}\approx10^{-9}\text{ yr}^{-1}.
\end{equation}

\noindent This number is several orders of magnitude lower than the observed GRB rate \citep{pod04}, and, as has been argued, such a PBH would be visible long before reaching any black hole.

Putting the unlikelihood of capture aside, one other comment is warranted concerning perhaps the most egregious assumption made by Barco. When specifying the constants of motion for the infalling PBH, Barco arbitrarily chooses ${L_s\approx2.23\times10^{-13}}$ and ${E_s=1.1\times10^{17}}$. The choice of $L_s$ is not too important---for a nonrelativistic object to fall within a radius $R$, it must have
\begin{equation}\label{eq:Ls}
    L_s^2<\frac{2(E_s+1)}{1-2M_\bullet/R}
\end{equation}

\noindent\citep{kou08}, which for the capture radius at the ISCO yields values of $L_s$ centered around 0 with a maximum close to unity. Within this range, different choices for $L_s$ have a negligible effect on the calculation of $T_\text{eff}$. However, $E_s$ is non-negligible, and Barco's value is much too high. Barco gives the justification that the central black hole will boost the emitter up to ultrarelativistic speeds, but his chosen value is simply absurd, for two reasons.

First, the specific energy is not ``boosted'' to high values as the PBH approaches the black hole (again, black holes don't have that kind of sucking power). $E_s$ is a constant of motion that does not change as the emitter speeds up close to the horizon. Even though a distant observer may see the emitter speed up in the black hole's gravitational well, the effect is exactly cancelled by gravitational time dilation so that the observed specific energy remains conserved.

The confusion may come from the fact that the local specific energy ${\gamma=u^t_\text{e}}$ given by equation~(\ref{eq:ue}) (or the equivalent equation~(1) in \citet{bar21}) does diverge as the emitter approaches the horizon. However, $\gamma$ measures the energy in the unphysical frame of ``shell observers'' who are able to remain static arbitrarily close to the horizon. In the physical frame of observers at rest at infinity, no such extreme boosting occurs---the fastest observed black hole inflow is only ${v\sim0.3c}$ \citep{pou18}, and an object in radial free fall from rest at infinity in a Schwarzschild spacetime will only reach a maximum observed speed of ${v\approx0.385c}$ (this is left as an exercise to the reader).

Second, now that it is established that black holes themselves will not substantially boost the observed energy of an infaller, all that remains is to show that the choice of the PBH's \emph{initial} specific energy is too high. Disregarding the effects of spacetime curvature, an object with a specific energy (and therefore Lorentz factor) of ${E_s=1.1\times10^{17}}$ would be so fast that it would be able to traverse the entire Milky Way galaxy in a mere .029 milliseconds of its own proper time. Its velocity would be
\begin{equation}
    v_0=0.999\ 999\ 999\ 999\ 999\ 999\ 999\ 999\ 999\ 999\ 99996c,
\end{equation}

\noindent which, given Barco's quoted mass of ${2.5\times10^{-13}M_\odot}$, would correspond to a total energy of ${3\times10^{61}}$ GeV. Not only is this number so far above the grand unified scale that no known laws of physics would apply (certainly not the classical general relativistic calculations done here), it also is only a few orders of magnitude away from the total estimated mass-energy content of the entire observable Universe.

What would be a more realistic choice for the value of $E_s$? The simplest option would be ${E_s=1}$, corresponding to a free-faller initially at rest at infinity (the initial velocity $v_0$ at infinity is related to the specific energy via the formula ${E_s=[1-(v_0/c)^2]^{-1/2}}$). Such an assumption was made in the context of PBHs in e.g.~\citet{jac73}.

If one wanted a more precise value for $v_0$ (or equivalently, $E_s$), one could follow Hawking's original supposition that $v_0$ lies in the range 50--1000 km/s (or equivalently, ${1.00000001<E_s<1.000006}$), similar to that of other bodies like stars and galaxies that move through the Universe \citep{haw71,bir16,yal21}. This range of values is not only consistent with cosmological constraints on the baryon-dark matter relative velocity after kinematic decoupling when PBHs would first be formed \citep{ali17,dvo14}, but it is also consistent with local observational measurements of the current velocity distribution of dark matter \citep{her18}. Some authors have supposed PBH velocities up to $\sim$0.6$c$ (${E_s\sim1.25}$) when studying gravitational interactions with other bodies like neutron stars \citep{cap13}, but even this value is close enough to ${E_s=1}$ that the simplified limit can be taken without question. Any other studies that consider higher, ultrarelativistic black hole velocities \citep[e.g.][]{dea78} only do so in a highly theoretical context and make no claims of any connections to PBHs or even to any physically realistic observations.

\section{Discussion}\label{sec:discussion}

Throughout the years since PBHs were first proposed, they have been used to abduce (i.e.\ back-explain) a variety of exotic phenomena, many of which have later been proven to be manifestly false. Even Hawking's original work on PBHs supposed not only that they could explain Weber's controversial claims of gravitational wave detections from 1970 but also that they could solve the solar neutrino problem that was later resolved with the theory of flavor oscillations \citep{haw71}. It would appear that such a false abduction has occurred once again in the context of thermal-dominated GRBs.

Three key premises of the arguments of \citet{bar21} give rise to the misleading assertion that PBH mergers with larger black holes produce observable GRBs: first, Barco makes special relativistic approximations that do not apply in the near-horizon regime; second, Barco assumes that an infalling emitter's radiation will be boosted and enhanced into a collimated beam when in fact it will be dimmed and redshifted from the observer's perspective; and third, Barco chooses a value for the PBH's initial velocity that is wildly, unrealistically high. Regarding the second point, from personal correspondences with Barco it is clear that the intent of that work was not to claim that radiation from PBHs would be Lorentz-boosted or enhanced directly because of the massive black hole; nevertheless, the claims as they are currently written in his paper are potentially misleading in that regard.

In summary, PBH-BH mergers cannot produce observable GRBs. Not only would their radiation be much too low to be detected, but even if it were high enough, the cooldown signature would look nothing like a GRB. Since the central black hole only acts to dim the PBH's radiation, the PBH would have been visible long before nearing another black hole's horizon. Instead of observing a burst, one would detect a constant streak, in the same way that meteors are observed for the entire duration of their descent through the sky, not just in the final milliseconds of their ablation.

As mentioned in Sec.~\ref{sec:introduction}, thermal-dominated GRBs already have well-known progenitors to explain their origins. These events are always extra-galactic and are associated with galaxies with rapid star formation (in contrast, a PBH origin would predict a higher occurrence rate in quenched galaxies, which host more black holes). The prevailing view, supported by multi-messenger observations, is that short GRBs are produced from kilonovae, and long GRBs are produced from collapsars. The only questions that remain are associated with how exactly the energy from these powerful events can be converted into the observed radiation through known physical processes.

\section*{Acknowledgements}

The author would like to thank Andrew Hamilton, Lia Hankla, and others in JILA for helpful discussions surrounding this work. The author would also like to thank the peer reviewers of this work for their comments and feedback.

\section*{Data Availability}

All data and numerical results generated or analysed during this study are included in this paper or the references therein. The author is happy to provide the \textsc{Mathematica} notebook used to produce these results to any interested readers.

\bibliographystyle{mnras}
\bibliography{biblio}

\begin{thebibliography}{}
\makeatletter
\relax
\def\mn@urlcharsother{\let\do\@makeother \do\$\do\&\do\#\do\^\do\_\do\%\do\~}
\def\mn@doi{\begingroup\mn@urlcharsother \@ifnextchar [ {\mn@doi@}
  {\mn@doi@[]}}
\def\mn@doi@[#1]#2{\def\@tempa{#1}\ifx\@tempa\@empty \href
  {http://dx.doi.org/#2} {doi:#2}\else \href {http://dx.doi.org/#2} {#1}\fi
  \endgroup}
\def\mn@eprint#1#2{\mn@eprint@#1:#2::\@nil}
\def\mn@eprint@arXiv#1{\href {http://arxiv.org/abs/#1} {{\tt arXiv:#1}}}
\def\mn@eprint@dblp#1{\href {http://dblp.uni-trier.de/rec/bibtex/#1.xml}
  {dblp:#1}}
\def\mn@eprint@#1:#2:#3:#4\@nil{\def\@tempa {#1}\def\@tempb {#2}\def\@tempc
  {#3}\ifx \@tempc \@empty \let \@tempc \@tempb \let \@tempb \@tempa \fi \ifx
  \@tempb \@empty \def\@tempb {arXiv}\fi \@ifundefined
  {mn@eprint@\@tempb}{\@tempb:\@tempc}{\expandafter \expandafter \csname
  mn@eprint@\@tempb\endcsname \expandafter{\@tempc}}}

\bibitem[\protect\citeauthoryear{{Abbott} et~al.,}{{Abbott}
  et~al.}{2017}]{abb17}
{Abbott} B.~P.,  et~al., 2017, \mn@doi [\prl] {10.1103/PhysRevLett.119.161101},
  \href {https://ui.adsabs.harvard.edu/abs/2017PhRvL.119p1101A} {119, 161101}

\bibitem[\protect\citeauthoryear{Ali-Ha\"{\i}moud \&
  Kamionkowski}{Ali-Ha\"{\i}moud \& Kamionkowski}{2017}]{ali17}
Ali-Ha\"{\i}moud Y.,  Kamionkowski M.,  2017, \mn@doi [Phys. Rev. D]
  {10.1103/PhysRevD.95.043534}, 95, 043534

\bibitem[\protect\citeauthoryear{Barco}{Barco}{2021}]{bar21}
Barco O.,  2021, \mn@doi [\mnras] {10.1093/mnras/stab1747}, 506, 806

\bibitem[\protect\citeauthoryear{{Berger}, {Fong}  \& {Chornock}}{{Berger}
  et~al.}{2013}]{ber13}
{Berger} E.,  {Fong} W.,   {Chornock} R.,  2013, \mn@doi [\apjl]
  {10.1088/2041-8205/774/2/L23}, \href
  {https://ui.adsabs.harvard.edu/abs/2013ApJ...774L..23B} {774, L23}

\bibitem[\protect\citeauthoryear{Bird, Cholis, Mu\~noz, Ali-Ha\"{\i}moud,
  Kamionkowski, Kovetz, Raccanelli  \& Riess}{Bird et~al.}{2016}]{bir16}
Bird S.,  Cholis I.,  Mu\~noz J.~B.,  Ali-Ha\"{\i}moud Y.,  Kamionkowski M.,
  Kovetz E.~D.,  Raccanelli A.,   Riess A.~G.,  2016, \mn@doi [Phys. Rev.
  Lett.] {10.1103/PhysRevLett.116.201301}, 116, 201301

\bibitem[\protect\citeauthoryear{Capela, Pshirkov  \& Tinyakov}{Capela
  et~al.}{2013}]{cap13}
Capela F.,  Pshirkov M.,   Tinyakov P.,  2013, \mn@doi [Phys. Rev. D]
  {10.1103/PhysRevD.87.123524}, 87, 123524

\bibitem[\protect\citeauthoryear{D'Eath}{D'Eath}{1978}]{dea78}
D'Eath P.~D.,  1978, \mn@doi [Phys. Rev. D] {10.1103/PhysRevD.18.990}, 18, 990

\bibitem[\protect\citeauthoryear{Dvorkin, Blum  \& Kamionkowski}{Dvorkin
  et~al.}{2014}]{dvo14}
Dvorkin C.,  Blum K.,   Kamionkowski M.,  2014, \mn@doi [Phys. Rev. D]
  {10.1103/PhysRevD.89.023519}, 89, 023519

\bibitem[\protect\citeauthoryear{Ghirlanda, Pescalli  \& Ghisellini}{Ghirlanda
  et~al.}{2013}]{ghi13}
Ghirlanda G.,  Pescalli A.,   Ghisellini G.,  2013, \mn@doi [\mnras]
  {10.1093/mnras/stt681}, 432, 3237

\bibitem[\protect\citeauthoryear{Hawking}{Hawking}{1971}]{haw71}
Hawking S.,  1971, \mn@doi [\mnras] {10.1093/mnras/152.1.75}, 152, 75

\bibitem[\protect\citeauthoryear{Henry, Feduniak, Silver  \& Peterson}{Henry
  et~al.}{1968}]{hen68}
Henry G.~R.,  Feduniak R.~B.,  Silver J.~E.,   Peterson M.~A.,  1968, \mn@doi
  [Phys. Rev.] {10.1103/PhysRev.176.1451}, 176, 1451

\bibitem[\protect\citeauthoryear{Herzog-Arbeitman, Lisanti, Madau  \&
  Necib}{Herzog-Arbeitman et~al.}{2018}]{her18}
Herzog-Arbeitman J.,  Lisanti M.,  Madau P.,   Necib L.,  2018, \mn@doi [Phys.
  Rev. Lett.] {10.1103/PhysRevLett.120.041102}, 120, 041102

\bibitem[\protect\citeauthoryear{Jackson \& Ryan}{Jackson \&
  Ryan}{1973}]{jac73}
Jackson A.~A.,  Ryan M.~P.,  1973, \mn@doi [\nat] {10.1038/245088a0}, 245, 88

\bibitem[\protect\citeauthoryear{Kouvaris}{Kouvaris}{2008}]{kou08}
Kouvaris C.,  2008, \mn@doi [Phys. Rev. D] {10.1103/PhysRevD.77.023006}, 77,
  023006

\bibitem[\protect\citeauthoryear{{Kulkarni} et~al.,}{{Kulkarni}
  et~al.}{1998}]{kul98}
{Kulkarni} S.~R.,  et~al., 1998, \mn@doi [\nat] {10.1038/27139}, \href
  {https://ui.adsabs.harvard.edu/abs/1998Natur.395..663K} {395, 663}

\bibitem[\protect\citeauthoryear{{MacFadyen}, {Woosley}  \&
  {Heger}}{{MacFadyen} et~al.}{2001}]{mac01}
{MacFadyen} A.~I.,  {Woosley} S.~E.,   {Heger} A.,  2001, \mn@doi [\apj]
  {10.1086/319698}, \href
  {https://ui.adsabs.harvard.edu/abs/2001ApJ...550..410M} {550, 410}

\bibitem[\protect\citeauthoryear{{Mazzali} et~al.,}{{Mazzali}
  et~al.}{2003}]{pao03}
{Mazzali} P.~A.,  et~al., 2003, \mn@doi [\apjl] {10.1086/381259}, \href
  {https://ui.adsabs.harvard.edu/abs/2003ApJ...599L..95M} {599, L95}

\bibitem[\protect\citeauthoryear{{Metzger} et~al.,}{{Metzger}
  et~al.}{2010}]{met10}
{Metzger} B.~D.,  et~al., 2010, \mn@doi [\mnras]
  {10.1111/j.1365-2966.2010.16864.x}, \href
  {https://ui.adsabs.harvard.edu/abs/2010MNRAS.406.2650M} {406, 2650}

\bibitem[\protect\citeauthoryear{{Misner}, {Thorne}  \& {Wheeler}}{{Misner}
  et~al.}{1973}]{mis73}
{Misner} C.~W.,  {Thorne} K.~S.,   {Wheeler} J.~A.,  1973, {Gravitation}.
W.~H. Freeman and Company

\bibitem[\protect\citeauthoryear{Nakar}{Nakar}{2007}]{nak07}
Nakar E.,  2007, \mn@doi [Phys. Rep.]
  {https://doi.org/10.1016/j.physrep.2007.02.005}, 442, 166

\bibitem[\protect\citeauthoryear{{Podsiadlowski}, {Mazzali}, {Nomoto},
  {Lazzati}  \& {Cappellaro}}{{Podsiadlowski} et~al.}{2004}]{pod04}
{Podsiadlowski} P.,  {Mazzali} P.~A.,  {Nomoto} K.,  {Lazzati} D.,
  {Cappellaro} E.,  2004, \mn@doi [\apjl] {10.1086/421347}, \href
  {https://ui.adsabs.harvard.edu/abs/2004ApJ...607L..17P} {607, L17}

\bibitem[\protect\citeauthoryear{Pounds, Nixon, Lobban  \& King}{Pounds
  et~al.}{2018}]{pou18}
Pounds K.~A.,  Nixon C.~J.,  Lobban A.,   King A.~R.,  2018, \mn@doi [\mnras]
  {10.1093/mnras/sty2359}, 481, 1832

\bibitem[\protect\citeauthoryear{Read}{Read}{2014}]{rea14}
Read J.~I.,  2014, \mn@doi [J. Phys. G: Nucl. Part. Phys.]
  {10.1088/0954-3899/41/6/063101}, 41, 063101

\bibitem[\protect\citeauthoryear{{Rybicki} \& {Lightman}}{{Rybicki} \&
  {Lightman}}{1985}]{ryb85}
{Rybicki} G.~B.,  {Lightman} A.~P.,  1985, Fundamentals of Radiative Transfer.
John Wiley \& Sons, Ltd, pp 1--50, \mn@doi{10.1002/9783527618170.ch1}

\bibitem[\protect\citeauthoryear{Ryde}{Ryde}{2004}]{ryd04}
Ryde F.,  2004, \mn@doi [\apj] {10.1086/423782}, 614, 827

\bibitem[\protect\citeauthoryear{{Sollerman} et~al.,}{{Sollerman}
  et~al.}{2006}]{sol06}
{Sollerman} J.,  et~al., 2006, \mn@doi [\aap] {10.1051/0004-6361:20065226},
  \href {https://ui.adsabs.harvard.edu/abs/2006A&A...454..503S} {454, 503}

\bibitem[\protect\citeauthoryear{{Tanvir}, {Levan}, {Fruchter}, {Hjorth},
  {Hounsell}, {Wiersema}  \& {Tunnicliffe}}{{Tanvir} et~al.}{2013}]{tan13}
{Tanvir} N.~R.,  {Levan} A.~J.,  {Fruchter} A.~S.,  {Hjorth} J.,  {Hounsell}
  R.~A.,  {Wiersema} K.,   {Tunnicliffe} R.~L.,  2013, \mn@doi [\nat]
  {10.1038/nature12505}, \href
  {https://ui.adsabs.harvard.edu/abs/2013Natur.500..547T} {500, 547}

\bibitem[\protect\citeauthoryear{Villanueva-Domingo, Mena  \&
  Palomares-Ruiz}{Villanueva-Domingo et~al.}{2021}]{vil21}
Villanueva-Domingo P.,  Mena O.,   Palomares-Ruiz S.,  2021, \mn@doi [Front.
  Astron. Space Sci.] {10.3389/fspas.2021.681084}, 8, 87

\bibitem[\protect\citeauthoryear{{Woosley} \& {Bloom}}{{Woosley} \&
  {Bloom}}{2006}]{woo06}
{Woosley} S.~E.,  {Bloom} J.~S.,  2006, \mn@doi [\araa]
  {10.1146/annurev.astro.43.072103.150558}, \href
  {https://ui.adsabs.harvard.edu/abs/2006ARA&A..44..507W} {44, 507}

\bibitem[\protect\citeauthoryear{Yalinewich \& Caplan}{Yalinewich \&
  Caplan}{2021}]{yal21}
Yalinewich A.,  Caplan M.~E.,  2021, \mn@doi [MNRAS: Lett.]
  {10.1093/mnrasl/slab063}, 505, L115

\bibitem[\protect\citeauthoryear{Yoshino, Takahashi  \& Nakao}{Yoshino
  et~al.}{2019}]{yos19}
Yoshino H.,  Takahashi K.,   Nakao K.-i.,  2019, \mn@doi [Phys. Rev. D]
  {10.1103/PhysRevD.100.084062}, 100, 084062

\makeatother
\end{thebibliography}

\bsp	
\label{lastpage}
\end{document}